# Anomalous Nernst effect in perovskite $La_{0.5}Ca_{0.5}CoO_3$

A. Ghosh, M. Manikandan, and R. Mahendiran

Department of Physics, National University of Singapore, 2 Science Drive 3, Singapore 117551, Republic of Singapore

**We report the occurrence of the anomalous Nernst effect (ANE) in polycrystalline perovskite $La_{0.5}Ca_{0.5}CoO_3$. The sample is ferromagnetic below $T_C$ = 147 K and resistivity shows non-metallic behavior above and below the $T_C$ with only a small negative magnetoresistance (~2%) around $T_C$. Field dependence of magnetization at 10 K shows large hysteresis with a coercive field of 6 kOe but a small magnetization ~ 0.64 $\mu_B$/Co even in a field of 50 kOe, which indicates the presence of magnetically heterogeneous ground state consisting of ferromagnetic and non-ferromagnetic phases. The field dependence of the Nernst thermopower ($S_{xy}$) at low temperatures shows complete saturation but the magnetization does not. This indicates that the ANE in $La_{0.5}Ca_{0.5}CoO_3$ depends only on the transport properties of the ferromagnetic phase, while it is not affected by the non-ferromagnetic phase. Due to the higher value of remnant $S_{xy}$, the magnetized polycrystalline sample exhibits ANE in absence of an external magnetic field.**

*Index Terms— Anomalous Nernst effect, Perovskite Oxide, Spincaloritronics.*

## I. INTRODUCTION

THE conversion of heat into spin-dependent electrical signals (spincaloritronics) has drawn immense attention in the last decade to find better, faster and energy saving technologies [1]–[6]. In this aspect, phenomena like, spin Seebeck effect (SSE) [5], anomalous Nernst effect (ANE) [3] and spin Nernst effect (SNE) [6] are being studied. The ANE has become very popular for its comparatively larger efficiency in converting heat into spintronic signal and simplicity in device fabrication [3], [4]. The Nernst effect is measured as the electric field developed in a material in y-direction due to the Lorentz force and/or spin-orbit coupling (SOC) experienced because of the mutual effects of thermal gradient ($\nabla T$) and magnetic field ($H$) applied independently in the $x$ and $z$-directions. The ordinary Nernst effect (NE) arises due the Lorentz force experienced by the conduction electrons of a material and thus it is proportional to the applied magnetic field. The ANE is deeply connected to SOC and its magnetic field dependence at a fixed temperature usually follows the magnetization curve of a material. However, the magnitude of the ANE thermopower has no correlation with the saturation magnetization [7]–[9].

To date, several metals, alloys and oxide materials have been investigated to understand the origin of ANE and to find ways to enhance its magnitude for better application possibilities. In addition to soft ferromagnets (Co, Ni, Fe and their alloys [2], [10]), ANE was also reported in hard ferromagnet such as $SmCo_5$ [11] and topological ferromagnets such as $Co_2MnGa$ [12], $Co_3Sn_2S_2$ [13], $Nd_2Mo_2O_7$ [14], etc. While $Co_2MnGa$ shows the highest anomalous Nernst thermopower ($S_{xy}^{ANE}$) around room temperature (~ 6 μV/K) [12], $SmCo_5$ exhibits the largest anomalous transverse thermoelectric conductivity ($\alpha_{xy}^{ANE}$) at room temperature (~ 4.6 A/K m) [11]. ANE is expected to be negligible in antiferromagnets. However, some antiferromagnets with non-colinear spin structure such as $Mn_3Sn$ [15], [16], $YbMnBi_2$ [17], $CoNb_3S_6$ [18], shows an unusually high value of ANE that does not scale with magnetization. The $\alpha_{xy}^{ANE}$ in $YbMnBi_2$ reaches ~ 10 μV/K m for $YbMnBi_2$ below 100 K, which is the largest value reported to date [17]. It is believed that Berry curvature of the conduction band near the Fermi level in these materials acts as fictious magnetic field to asymmetrically scatter spin up and spin down charge carries leading to the large ANE values. In this context, ANE in oxide materials are relatively less explored. $La_{0.7}Sr_{0.3}CoO_3$ [2], [19], $Pr_{0.5}Sr_{0.5}CoO_3$ [20], $R_{0.6}Sr_{0.4}CoO_3$ (R = La, Pr, Nd) [21] and $Fe_3O_4$ [1] are few of these oxides that have shown their potential for ANE in both the single and polycrystalline phases.

While $LaCoO_3$ is a non-magnetic insulator, divalent alkaline earth cation ($Sr^{2+}$, $Ca^{2+}$, $Ba^{2+}$) substitution for the $La^{3+}$ rare-earth ion brings in ferromagnetism and metallic conduction. The divalent cation substitution not only partially transforms $Co^{3+}$ into $Co^{4+}$, but also transforms the spin state. While $Co^{3+}$ ion is primarily in low spin state ($t_{2g}^6e_g^0$) in $LaCoO_3$, it is transformed into intermediate spin state ($t_{2g}^5e_g^1$) in the doped compounds. It is believed that ferromagnetism in these oxides is due to double exchange interaction between intermediate spins $Co^{3+}$ ($t_{2g}^5e_g^1$) and low spins $Co^{4+}$ ($t_{2g}^5e_g^0$) ions [22]. $La_{0.5}Ca_{0.5}CoO_3$ possesses an orthorhombic (space group: $Pnma$) structure and shows a ferromagnetic (FM) ordering near 150 K [23]. It was suggested that the low temperature phase may be heterogeneous in which the non-percolating FM phase may coexist with the non-FM phase [24], [25], and thus it may be interesting to study the ANE. In this work, we studied the ANE in polycrystalline $La_{0.5}Ca_{0.5}CoO_3$ and found that the ANE is mainly sensitive to the FM part of the sample and thus, does not follow the non-saturating magnetization as observed due to the alignment of the non-FM phase at low temperatures. The sample also exhibited zero-field ANE in the magnetized state.

## II. EXPERIMENTAL

Polycrystalline $La_{0.7}Ca_{0.3}CoO_3$ was synthesized by the solid-state reaction from stoichiometrically mixed pre-heated powders of $La_2O_3$, $CaCO_3$, and $Co_3O_4$. After consecutive grinding and heating of the powder several times at 1000ºC and 1100º C for 12 hours, the powder was finally pressed into pellets and sintered at 1200ºC for 24 hours in the air. Magnetic measurements were carried out in a vibrating sample magnetometer (VSM) attached to the physical property measurement system (PPMS). Resistivity and longitudinal thermopower were simultaneously measured in a homemade probe attached to the PPMS. For ANE measurement, a rectangular-shaped sample of typical dimension ~ 5 × 3 × 0.5 mm³ was cut and polished. The ANE was measured in a custom designed homemade probe attached to the PPMS which is discussed in detail in our earlier work [20]. In this setup, the sample is placed between the top of two copper blocks maintained at two different temperatures creating an in-plane temperature difference ($\Delta T$). The bottom and top surfaces of the sample may have different temperatures depending on the thermal conductivity of the measured sample and thus it can create an error in measuring the actual value of $\Delta T$ [26]. This can underestimate the obtained value of Nernst thermopower ($S_{xy}$) by 5-10% for the sample reported here. The $S_{xy}$ was measured as a function of applied magnetic field ($H$), and the voltages generated from longitudinal thermopowers of the sample and measurement leads were subtracted as, $S_{xy} = \frac{1}{2} \frac{(V_{xy}(+H) - V_{xy}(-H))/l_y}{\Delta T_x/l_x}$ where $l_y$ is the distance between the measurement leads along $y$-direction and $l_x$ is the distance between the two copper blocks along the $x$-direction (a schematic is given in Fig. 3(d) showing the measurement configuration).

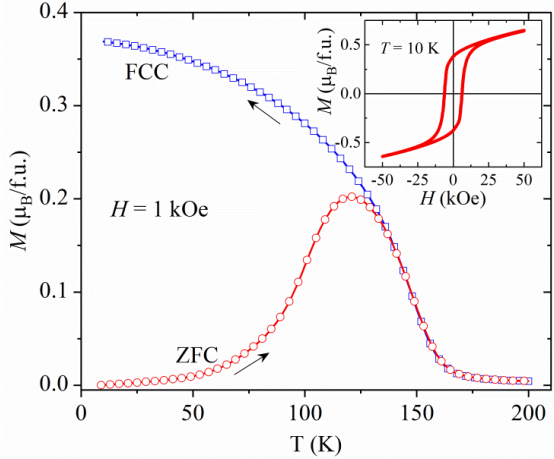

Fig. 1. Zero field cooled (ZFC) and field cooled cooling (FCC) temperature dependence of the magnetization at $H$ = 1 kOe. Inset: Field dependence of the magnetization measured at $T$ = 10 K.

## III.  RESULT AND DISCUSSION

### A.  Magnetic Properties

Fig. 1 shows the temperature dependence of zero-field cooled (ZFC) and field cooled cooling (FCC) magnetization for La$_{0.5}$Ca$_{0.5}$CoO$_3$ measured in the presence of $H$ = 1 kOe magnetic field. ZFC magnetization increases with temperature, exhibit a maximum near 120 K and decreases rapidly across the Currie temperature ($T_C$) ~ 147 K. The FCC curve follows the ZFC magnetization near the $T_C$, bifurcate near 120 K and increase monotonically with decreasing the temperature. Such irreversibility in magnetization may arise due to the breakdown of long-range ferromagnetic order and emergence of cluster spin-glass state or freezing of magnetic domains in random directions in the ZFC mode if the measuring field is much lower than the anisotropy field [27], [28]. The field dependence of magnetization ($M$-$H$ curve) at 10 K plotted in the inset of Fig. 1 exhibits hysteresis with a large coercive field, $H_C$ ~ 6 kOe. The $M$-$H$ curve at 10 K (as shown in the inset) does not show saturation up to 50 kOe, where $M$ reaches 0.65 $\mu_B$/Co, a value much less than 1.62 $\mu_B$/Co found in La$_{0.5}$Sr$_{0.5}$CoO$_3$ [22]. However, the observed value of magnetization in our composition is closer to 0.61 $\mu_B$/Co at 10 K as obtained by neutron diffraction study of La$_{0.9}$Ca$_{0.1}$CoO$_3$ sample, in which long range ferromagnetism was already found [29]. It is likely that the FM phase coexists with the non-FM phase in our sample. While Co³⁺/Co⁴⁺ ions are in intermediate spin / low-spin configurations mediating FM interaction in the FM phase, Co³⁺ ions are in low-spin state (S = 0) in the non-FM phase. However, we are not discussing this aspect in detail as the scope of this paper is restricted to study the ANE.

### B.  Longitudinal Resistivity and Thermopower

Fig. 2(a) shows the temperature dependence of the longitudinal resistivity ($\rho_{xx}$) measured at $H$ = 0 and 50 kOe fields. The sample exhibits non-metallic resistivity where the value of $\rho_{xx}$ increases more than 100 times from 250 K to 10 K. As the magnetic field has an insignificant effect on the resistivity in the scale of the temperature dependent plots, we measured the field sweep in the vicinity of $T_C$ and obtained ~ -2% magnetoresistance (MR) at 150 K as shown in the inset of Fig. 2(a). The MR was calculated as the percentage change in $\rho_{xx}$ under the magnetic field which can be expressed as, $MR = \frac{(\rho_{xx}^H - \rho_{xx}^0)}{\rho_{xx}^0} \times 100$ % where $\rho_{xx}^0$ and $\rho_{xx}^H$ are respectively the resistivity values in the absence and presence of an external magnetic field.

The temperature dependence of the longitudinal thermopower ($S_{xx}$) is plotted in Fig. 2(b) for the same fields as the resistivity. The positive sign of $S_{xx}$ indicates holes as the majority carrier. $S_{xx}$ shows a clear change in slope near the $T_C$ and continuously decreases at low temperatures. The change in slope of $S_{xx}$ around $T_C$ implies a decrease in scattering felt by charge carriers diffusing under the temperature gradient due to spin alignment, $i.e.$, decrease in spin-disorder scattering. The value of thermopower decreases in the presence of $H$ = 50 kOe in a narrow temperature region around $T_C$, whereas both the curves merge far below and above $T_C$. Application of the magnetic field enhances the spin alignment and thus reduces spin-disorder scattering near $T_C$ which leads to a longer mean free path for the heat carrying charges as compared to the zero field case. Since spin-disorder scattering is negligible even in zero field at temperatures much below $T_C$, the applied magnetic field has little or no influence on the value of thermopower as seen previously in other oxides [30], [31]. The magnetothermopower (MTEP) was estimated as the percentage change in $S_{xx}$ due to the presence of a magnetic field $\left(MTEP = \frac{(S_{xx}^H - S_{xx}^{H=0})}{S_{xx}^{H=0}} \times 100 \right.$ %$\left. \right)$, where $S_{xx}^{H=0}$ and $S_{xx}^H$ are the values of the longitudinal thermopower respectively in the absence and presence of an external magnetic field. At 150 K, a maximum MTEP of -7 % is obtained due to $H$ = 50 kO (Inset of Fig. 2(b)). While the thermopower is sensitive to dynamics of charges within grains, the electrical resistivity is affected by both grain and grain boundary resistances. In magnetic samples, it causes semiconducting like resistivity across ferromagnetic transition, but the thermopower is found to show a prominent anomaly only around the $T_C$ [32].

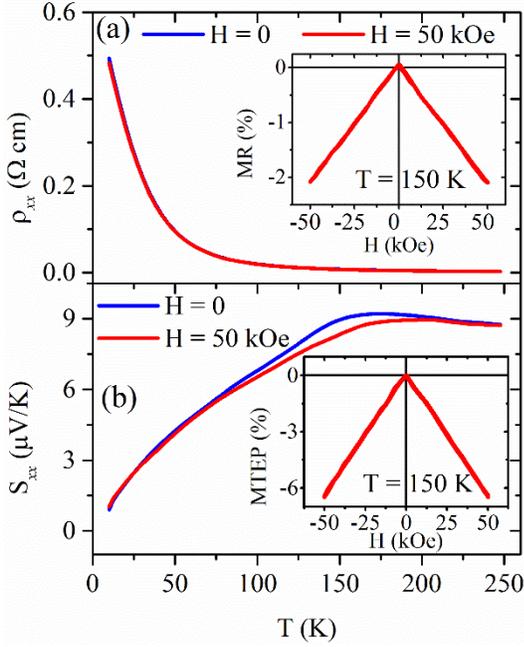

Fig. 2. (a) Temperature dependence of the longitudinal resistivity ($\rho_{xx}$) measured at $H = 0$ and 50 kOe. Inset: Field dependence of the magnetoresistance (MR) measured at 150 K. (b) Temperature dependence of the longitudinal thermopower ($S_{xx}$) measured at $H = 0$ and 50 kOe. Inset: Field dependence of the magnetothermopower (MTEP) measured at 150 K.

### C. Field Dependence of Magnetization and Anomalous Nernst Effect

Fig. 3(d) shows schematic diagram of the measurement configuration indicating the direction of applied $\Delta T$ along $x$, magnetic field along $z$ and measured voltage along $y$-directions Fig. 3(a) – 3(c) show the $M$-$H$ curves (right vertical axis) and field dependences of the $S_{xy}$ (left vertical axis) measured at 25, 120 and 200 K. Although, both $M(H)$ and $S_{xy}(H)$ show similar hysteresis, the data at 25 K indicates that while $M$ does not show saturation in the maximum available field, $S_{xy}$ saturates above $H = 20$ kOe. This indicates that the ANE in this material does not follow the total magnetization at 25 K. The total magnetization can be considered as the sum of the contributions coming from the FM and the non-FM phases. In this case, the nature of the $S_{xy}$ vs $H$ curve follows only the FM part and this implies that the non-FM contribution present in the sample does not contribute to the Nernst effect. However, $S_{xy}$ vs $H$ at 120 K shows non-saturating behavior and overlaps with the $M$ vs $H$ curve (Fig. 3(b)). At 200 K (Fig. 3(c)), both $S_{xy}$ and $M$ increase linearly with the magnetic field as only the normal Nernst effect is active in the paramagnetic state. Figs. 3(e) and (f) respectively show the $M$-$H$ and $S_{xy}$-$H$ curves measured at $T = 60, 80, 120$ and $140$ K in the low-field regime (-10 kOe $\leq H \leq$ +10 kOe) for clarity. Both the magnetization and $S_{xy}$ trace similar hysteresis loops with large $H_C$. The hysteresis becomes narrow with increasing temperature. The remnant value of $S_{xy}$ is significant in this material due to the larger value of $H_C$ which gives rise to the zero-field ANE discussed later in this paper.

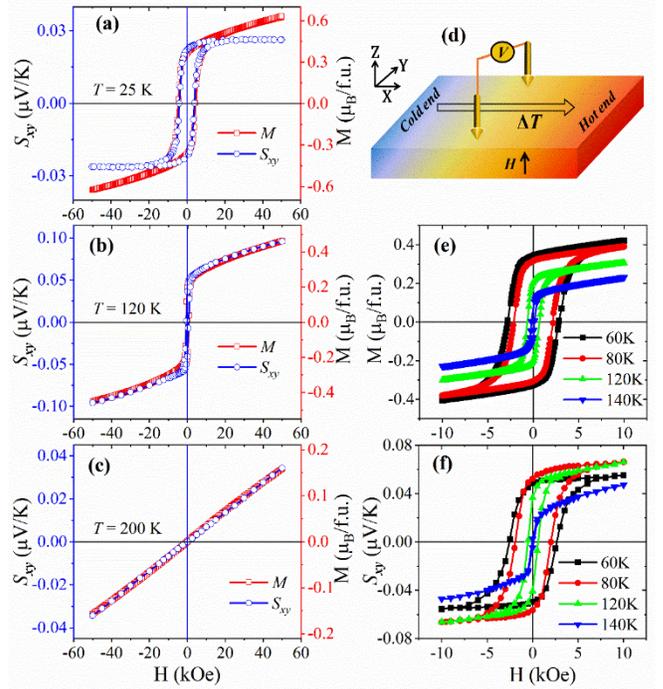

Fig. 3. Field dependences of magnetization (right vertical axis) and Nernst thermopower, $S_{xy}$ (left vertical axis) measured at (a) $T = 25$ K, (b) $T = 120$ K and (c) $T = 200$ K. (d) Schematic representation of the measurement configuration for the Nernst effect. Expanded view of the low field ( -10 kOe $\leq H \leq 10$ kOe) dependence of (e) magnetization and (f) $S_{xy}$ measured at $T = 60, 80, 120$ and $140$ K.

### D. Temperature Dependence of Anomalous Nernst Effect

The high field region ($H = 30$-$50$ kOe) of the $S_{xy}$ vs $H$ curves was fitted linearly to extract the value of $S_{xy}$ due to NE ($S_{xy}^{NE}$) and ANE ($S_{xy}^{ANE}$) obtained at different temperatures [19]. $S_{xy}^{NE}$ was obtained from the slope of the linear fit as; $S_{xy}^{NE}(H) = S_{xy}^{Slope}H$ and, the anomalous contribution was extracted as; $S_{xy}^{ANE}(H) = S_{xy}(H) - S_{xy}^{NE}(H)$. The extracted $S_{xy}^{ANE}$ is plotted as a function of temperature in Fig. 4(a) for $H = 5, 10$ and $50$ kOe. As the temperature decreases below 180 K, $S_{xy}^{ANE}$ increases rapidly due to magnetic ordering across the $T_C$, goes through a maximum near 100 K, and decreases monotonically towards zero at low temperatures. Although the $S_{xy}^{ANE}$ vs $T$ curves for $H = 10$ and $50$ kOe show a similar behavior at low temperatures, the same for $H = 5$ kOe decreases much rapidly due to non-saturation of the sample because of the increased $H_C$. In addition to that, the zero-field $S_{xy}^{ANE}$ is also plotted in the same figure (star symbol) which is same as the remnant value of $S_{xy}$ and obtained from the $S_{xy}$ vs $H$ curves at $H = 0$ while reducing the field from 50 kOe. It is interesting to note that even at $T = 120$ K, an adequate value of zero-field ANE signal is obtainable. This indicates that ANE in $La_{0.5}Ca_{0.5}CoO_3$ can be obtained by applying $\Delta T$ in the absence of $H$ if the sample is already magnetized.

Fig. 4(b) represents the temperature dependence of the anomalous transverse thermoelectric conductivity which can be expressed as [14],

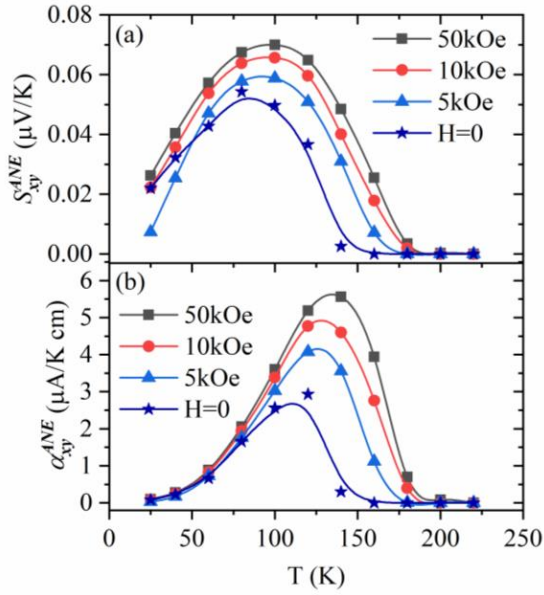

Fig. 4. Temperature dependence of the (a) anomalous Nernst thermopower ($S_{xy}^{ANE}$) and (b) anomalous transverse thermoelectric conductivity ($\alpha_{xy}^{ANE} \approx S_{xy}^{ANE}/\rho_{xx}$) obtained for $H = 0, 5, 10$ and 50 kOe fields. $S_{xy}^{ANE}$ and $\alpha_{xy}^{ANE}$ at $H = 0$ represent the remnant values recorded after reducing the field to 0 from 50 kOe. Data points are connected through spline fitting to guide the eye.

$$\alpha_{xy}^{ANE} = [S_{xy}^{ANE} - S_{xx}(\rho_{xy}/\rho_{xx})]/\rho_{xx} \qquad (1)$$

where $\rho_{xy}$ is the Hall resistivity. The reported value of the Hall angle, $\theta_H \approx \rho_{xy}/\rho_{xx}$ is ~$6.5 \times 10^{-5}$ at 100 K for La$_{0.85}$Ca$_{0.15}$CoO$_3$ single crystal [33]. This makes the second term of Eq. (1) negligible. Thus, for the samples with very low $\theta_H$, Eq. (1) can be simplified to $\alpha_{xy}^{ANE} \approx S_{xy}^{ANE}/\rho_{xx}$. The $\alpha_{xy}^{ANE}$ increases rapidly across the ferromagnetic ordering temperature, goes through a peak, and rapidly decreases towards zero on the low temperature side. In metallic systems, the $\alpha_{xy}^{ANE}$ follows a $T$-linear behavior at low temperatures obeying the Mott's rule [34], but a clear deviation from the linear behavior is observed in the present compound. The resistivity of the sample increases from ~ 0.02 $\Omega$ cm at 100 K to ~ 0.3 $\Omega$ cm at 25 K which in turn drastically diminishes the value of $\alpha_{xy}^{ANE}$ and thus such anomaly is observed.

## IV. CONCLUSION

In summary, we have studied the field and temperature dependence of the anomalous Nernst effect (ANE) in La$_{0.5}$Ca$_{0.5}$CoO$_3$. The ANE shows an abrupt increase near the ferromagnetic ordering, goes through a peak, and decreases to zero at low temperatures. While the ANE signal shows saturation at low temperatures, magnetization does not saturate. An interesting feature of this compound is the non-zero ANE value in zero magnetic field at cryogenic temperatures which might be useful for heat into spin-dependent thermoelectric conversion. Future work needs to address the magnetic ground state of La$_{0.5}$Ca$_{0.5}$CoO$_3$ by neutron diffraction and dynamic magnetic susceptibility.

## ACKNOWLEDGMENT

R.M. acknowledges the Ministry of Education, Singapore for supporting this work (Grant. No. R144-000-428-114 and R144-000-422-114).